\documentclass[12pt]{iopart}
\newcommand{\PeM}{{\rm PM}}
\newcommand{\MM}{{\rm MM}}
\newcommand{\Sfold}{{\rm Sfold}}
\newcommand{\Nfold}{{\rm Nfold}}
\newcommand{\Pfold}{{\rm Pfold}}
\newcommand{\SN}{{\rm SN}}
\newcommand{\SeS}{{\rm SS}}
\newcommand{\NN}{{\rm NN}}
\newcommand{\PeS}{{\rm PS}}
\newcommand{\PN}{{\rm PN}}
\newcommand{\Sum}{{\rm sum}}
\usepackage{graphicx,amssymb}% Include figure files
\usepackage{color}

\begin{document}

\title[Physico-chemical modelling of target depletion]{Physico-chemical modelling of target depletion during hybridisation on oligonulceotide microarrays}  

\author{Conrad J.\ Burden$^1$ and Hans Binder$^2$}
\address{$^1$Centre for Bioinformation Science\\
      Mathematical Sciences Institute \\
      Australian National University,
      Canberra, ACT 0200, Australia}
\ead{Conrad.Burden@anu.edu.au}

\address{$^2$Interdisciplinary Centre for Bioinformatics of Leipzig University\\
      D-4107 Leipzig \\
      Haertelstrasse 16-18,
      Germany}
\ead{binder@rz.uni-leipzig.de}

\begin{abstract}
The effect of target molecule depletion from the supernatant solution is incorporated into a physico-chemical model of hybridisation on oligonucleotide microarrays.  Two possible regimes are identified: local depletion, in which depletion by a given probe feature only affects that particular probe, and global depletion, in which all features responding to a given target species are affected.  Examples are given of two existing spike-in data sets experiencing measurable effects of target depletion.  The first of these, from an experiment by Suzuki et al.\ using custom built arrays with a broad range of probe lengths and mismatch positions, is verified to exhibit local and not global depletion.  The second dataset, the well known Affymetrix HGU133a latin square experiment is shown to be very well explained by a global depletion model.  It is shown that microarray calibrations relying on Langmuir isotherm models which ignore depletion effects will significantly underestimate specific target concentrations.  It is also shown that a combined analysis of perfect match and mismatch probe signals in terms of a simple graphical summary, namely the hook curve method, can discriminate between cases of local and global depletion.  
\end{abstract}


\maketitle

\section{Introduction}
\label{sec:Introduction}

Physico-chemical models describing the processes involved in converting concentrations of specific RNA or DNA targets hybridised onto oligonucleotide microarrays to observed fluorescence intensities have become commonplace~\cite{Hekstra03, Held03, Burden04, Binder04, Binder06a, Binder06b, Carlon06, Heim06, Burden06, Held06, Burden08, Mulders09, Binder09a, Nguyen09}.  The ultimate aim of such models has in general been to provide biologists with practical algorithms for estimating absolute specific target concentrations in the presence of a complex non-specific background from fluorescence intensity data.  Early models inspired by Langmuir adsorption theory, which applied standard physical chemistry to the hybridisation of specific and non-specific targets to the microarray surface, predicted a hyperbolic response function~\cite{Hekstra03, Held03} which has been verified with reasonable accuracy~\cite{Burden04} for the Affymetrix U95a Latin Square spike-in experiment~\cite{AffyLatinSq}.  Refinements of the model to include the effects of probe and target folding and bulk hybridisation in the supernatant solution~\cite{Binder06a, Carlon06} maintain the hyperbolic shape of the response function while decreasing the effective adsorption rate constant.  Including the effects of post-hybridisation washing~\cite{Burden06,Held06} also maintains the hyperbolic shape of the response function and is able to explain an asymptotic response in the limit of high target concentrations which is below full saturation of the probe feature and decreases with probe-target binding affinity~\cite{Burden08}.  

The above physico-chemical models generally assume that the concentration of target molecules in the supernatant solution is not appreciably depleted by the hybridisation reaction.  However, in order to explain their data from spike in experiments which run to very low spike-in concentratons~\cite{Suzuki07}, Ono et al.~\cite{Ono08} have recently extended the accepted adsorption model to include such target depletion effects.   Their model predicted an interesting saturation effect which was borne out by experiment.  As well as the usual saturation effect, in which the number of available probe molecules becomes exhausted in the limit of high target concentration for a fixed probe type, a second saturation effect occurs when the number of target molecules is exhausted in the limit of high binding affinity at fixed target concentration.  This limit was realised by including on a custom-built microarray a series of features of increasing probe length.  

In the current paper we extend the Ono model by identifying two types of target depletion, which we term ``local depletion'' and ``global depletion''.  By local depletion we mean that depletion of target molecules in the supernatant solution by a hybridisation to a given probe feature only affects that particular probe feature.  This is essentially Ono et al.'s ``finite hybridisation model''.  This regime is relevant when diffusion and/or convection of targets is slow compared with the hybridisation and probe features responsive to the same target species are spatially separated on the microarray.  By global depletion we mean that all probe features responding to a given target species are mutually affected by depletion of that species from the supernatant solution.   Global depletion is relevant for spatially separated features undergoing permanent agitation of the hybridisation solution, if equilibrium includes rapid diffusion of transcripts through the microarray cartridge, or for neighbouring features such as the perfect match/mismatch (PM/MM) probes on the older designs of Affymetrix GeneChips.  

We fit the models to two spike-in datasets. The first, that of Suzuki et al.~\cite{Suzuki07}, which covers a broad range of spike-in concentrations and probe lengths and for which we verify that the local model, not the global model, is relevant, is dealt with in Section~\ref{SuzukiDataAnalysis}.  The second, the U133A Affymetrix Latin Square data set~\cite{AffyLatinSq}, for which the global model is appropriate, is dealt with in Section~\ref{AffyDataAnalysis}.  For this data set we find that the global model of depletion entails a substantial improvement on earlier reported fits by a hyperbolic response function~\cite{Burden08}.  As well as fitting response functions or so-called ``isotherms'', we analyse the data sets in terms of the recently developed hook curve formalism~\cite{Binder08a, Binder08b} designed for the calibration of microarrays whose design includes PM/MM pairs. The hook curve method turns out to be a clear and easily implemented indicator of which depletion regime, local or global, is relevant to a particular dataset.  

Full details of our local and global depletion models, including specific and non-specific hybridisation of target molecules to probes at the microarray surface and of targets within the supernatant solution and the folding of target and probe molecules, are set out in \ref{sec:PCModel}.  Some technical details of the analysis of the global depletion model are given in \ref{sec:IsoShape}.  

Other than the work of Ono et al.\ and a related project~\cite{Nguyen09}, we are aware of only one other extensive attempt to incorporate target depletion from the supernatant solution during hybridisation into a physico-chemical model of microarrays, namely a recent publication by Li et al.~\cite{Li08}.  In Section~\ref{sec:CritEval} we give a critical evaluation pointing out a number of errors in the Li et al.\ model, with details given in \ref{sec:LiCritique}.  

\section{The Suzuki data set: an example of local depletion}
\label{SuzukiDataAnalysis}

Suzuki et al.~\cite{Suzuki07} have carried out experiments in which a set of 150 cDNA target sequences, with and without a complex background,  are hybridised onto custom arrays containing features with probes ranging in length from $\ell =$ 14 to 25 DNA bases.  The probe designs include perfect matches and mismatches, the mismatches being in each possible position ($1,\ldots,\ell)$ and of each possible nucleotide.  Spike-in concentrations covered a broad range from 1.4~fM to 1.4~nM.  The purpose of the experiment was to determine probe lengths and mismatch positions which optimise the discrimination between PM and MM signals.  Because the spike-in concentrations run to very low values, depletion cannot be ignored~\cite{Ono08}.  Of the two physico-chemical models described in \ref{sec:PCModel}, we demonstrate below that this data set is an example of local rather than global target depletion.  This result is reasonable:  The large set of PM and MM probes addressing any one target species must extend over distances large compared with the nearest neighbour distance on the chip.  The remaining question, which we settle below in favour of local depletion, is whether diffusion or convection of target molecules is slow (local depletion) or fast (global depletion) relative to the rate of hybridisation.  

\subsection{Theory}

For the case of local depletion, the coverage fraction $0 \le \theta \le 1$ of fluorescent dye carrying target molecules bound to a given probe feature at the microarray surface is shown in \ref{sec:ALocalDepletion} to be 
\begin{equation}
\theta = \frac{X_N + K_S \left(x_S - p \theta_S \right)}{1 + X_N + K_S \left(x_S - p \theta_S \right)} ,               \label{thetaSoln1}
\end{equation}
where $x_S$ is the spiked-in probe-specific target concentration, $p$ is an effective molar concentration of probe molecules immobilised on the microarray surface, $X_N$, called the non-specific binding strength, is a dimensionless measure of the degree of non-specfic binding and $K_S$ is an effective equilibrium constant for the binding of specific targets accounting for several chemical reactions including surface and bulk hybridisation and molecular folding.  The coverage fraction $\theta_S$ of specific targets only is given by Eq.~(\ref{thetaSfromQuad}).   The model of  \ref{sec:ALocalDepletion} also allows for the consideration of post-hybridisation washing, which is signalled by differing responses of PM and MM features to saturation target concentrations~\cite{Burden06}.  There seems to be little evidence that washing is significant for this dataset (see Figure~\ref{fig:SuzukiData}), and for convenience we set the washing survival factors to unity in the current analysis.  

\begin{figure}
\includegraphics[scale=0.75]{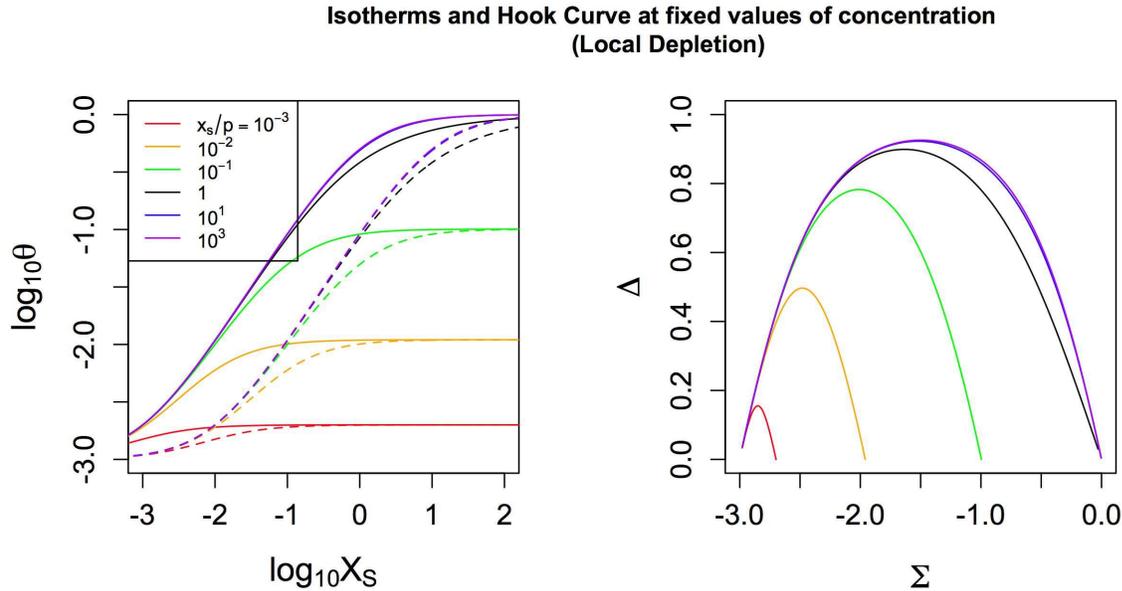}
\centering
\caption{Theoretical isotherms and hook curves derived from the local depletion model of \ref{sec:ALocalDepletion}.  Each curve represents the response of the coverage fraction $\theta$ to variations of the specific binding strength for the PM probes,  $X^\PeM_S = K^\PeM_S x_S$ at fixed specific target concentration $x_S$.  Since $\log K^\PeM_S \propto$ free binding energy of hybridisaton (see text), the horizontal scale in the isotherm plot can be thought of as a measure of probe length.  Input parameters are: $x_S/p$ as indicated by colour in the legend; $X_N = 10^{-3}$ and $K_S^\MM = 0.1 K_S^\PeM$.   Isotherms for PM probes are plotted as solid lines, and for MM probes as dashed liines.}
\label{fig:theoIsothermLocal}
\end{figure}

The log of the effective equilibrium constant $K_S$ is expected to be approximately proportional to probe length.  This follows from the definition Eq.~(\ref{KSdefLocal1}) and the relationship $K_\PeS \propto e^{\Delta G/(RT)}$ relating the hybridisation constant to free binding energy $\Delta G$, which is well approximated by the SantaLucia nearest neighbour stacking model~\cite{SantaLucia98}.  Ono et al.~\cite{Ono08} make use of this result to consider isotherms relating coverage fraction to probe length, which we reproduce from the theoretical local depletion model in the left panel of Figure~\ref{fig:theoIsothermLocal}.  In calculating these curves we use an assumption that the ratio $K_S^\PeM/K_S^\MM$ is independent of probe length.  This is justified since $K_S^\PeM/K_S^\MM \approx e^{\Delta\Delta G/(RT)}$ where $\Delta\Delta G = \Delta G^\PeM - \Delta G^\MM$ is, on average, independent of probe length by virtue of the nearest neighbour stacking model.  

Two saturation behaviours in the limit of large probe length, or high binding strength $X_S = K_S x_S$, are immediately apparent.  From Eqs.~(\ref{thetaN2}) and  (\ref{thetaSfromQuad}) one obtains
\begin{equation}
\lim_{X_S \rightarrow \infty} \theta = \lim_{K_S \rightarrow \infty} (\theta_S + \theta_N) = \left\{ 
\begin{array}{ll}
1 & \mbox{if } x \ge p, \\
x/p & \mbox{if } x < p. 
\end{array}  \right.
\end{equation}
In the case $x > p$, where the concentration of specific target exceeds the effective concentration of probes, the probes become saturated ($\theta = 1$) and any residual unbound targets remain in solution.  In the case $x < p$, where the probe concentration exceeds that of the targets, the free targets are completely depleted and the maximum fluorescence intensity decreases with decreasing target concentration ($\theta = x/p$).  Note also that for any PM/MM pair, the saturation intensity depends only on specific target concentration and not on the presence of mismatches.  

Also shown in the right panel of Figure~\ref{fig:theoIsothermLocal} are the predicted hook curves for varying probe length in the case of local depletion.  The hook curve method~\cite{Binder08a, Binder08b} was originally developed to analyse data from microarrays whose design includes PM/MM pairs, but can be applied to any pair of probe features addressing the same specific target.  The method processes the PM/MM intensities $I^\PeM$ and $I^\MM$ using the transformation 
\begin{equation}
\Delta = \log_{10} I^\PeM - \log_{10} I^\MM , \qquad
\Sigma = \frac{1}{2} \left\langle \log_{10} I^\PeM + \log_{10} I^\MM \right\rangle, \label{SigmaDeltaDef}
\end{equation}
where, for Affymetrix GeneChips, the angular brackets denote averaging over probes within a probeset.  Smoothing the $\Delta$ versus $\Sigma$ plot provides a hook curve, whose characteristic shape typically assumes the concave downwards curve shown in Figure~\ref{fig:theoIsothermLocal}.  

In previous implementations \cite{Binder08a,Binder08b} the hook curve has been considered as a trajectory in the $\Sigma$-$\Delta$ plane as the specific binding strength $X_S = K_S x_S$ varies due to changes in specific target concentration $x_S$ while the binding affinity $K_S$ is held fixed.  For the Suzuki data set we use a {\em different} and more appropriate implementation which specifically exploits the broad range of binding affinities arising from probe lengths which vary from 14 to 25 mer.  That is, Figure~\ref{fig:theoIsothermLocal} plots the hook curve as a trajectory traced out by varying binding affinity $K_S$ at fixed values of target concentration $x_S$.  The left hand end of the hook curve ($X_S =0$) is determined by non-specific hybridisation and will not vary significantly with probe length.  The right hand end of the hook curve ($X_S \rightarrow \infty$) is determined by the saturation intensity, and is expected to shift leftwards for subcritical specific target concentrations $x < p$.  

\begin{figure}
\includegraphics[scale=0.75]{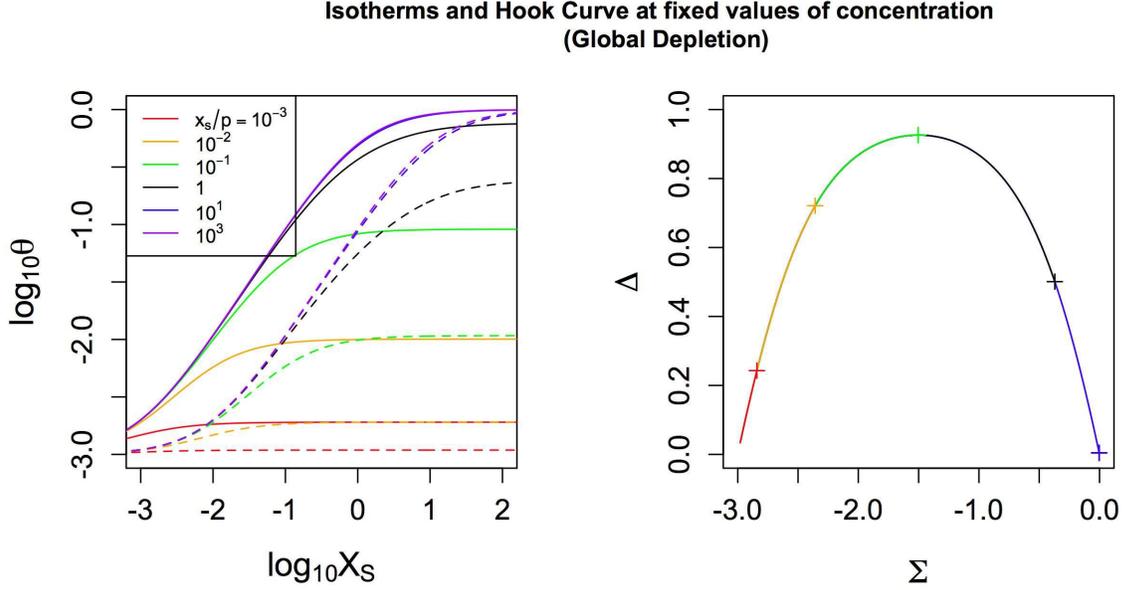}
\centering
\caption{Theoretical isotherms and hook curves derived from the global depletion model of \ref{sec:AGlobalDepletion}.  The input parameters and curve conventions are the same as for Figure~\ref{fig:theoIsothermLocal}.  Also shown in the right panel are the right hand end points of individual hook curves indicated by a $+$ sign.  As explained in the text, the shape of the hook curve remains unchanged as $x_S$ varies, except that the curve terminates at different right hand points at subcritical concentrations.  Hook curves for all values of $x_S/p$ start at the same left hand point.  Note that the critical value of specific concentration, below which the isotherm saturates at $\theta < 1$ now occurs at a value $x_{\rm crit}/p > 1$ as the depleted targets are shared among more than one probe feature.}
\label{fig:theoIsothermGlobal}
\end{figure}

By contrast, the theoretical isotherms and hook curves for the case of global depletion are shown in Figure~\ref{fig:theoIsothermGlobal}.  For a given probe feature $P$, the theoretical isotherm, derived in \ref{sec:AGlobalDepletion}, is now (see Eq.~(\ref{thetaSolnP}))
\begin{equation}
\theta^P = \frac{X_N^P + K_S^P (x_S - p \theta_\Sum)}{1 + X_N^P + K_S^P (x_S - p \theta_\Sum)}, 
                \label{thetaSolnGlobal}
\end{equation}
where $ \theta_\Sum = \sum_P \theta^P$ is the sum total of specific target coverage fractions over all probe features addressing the relevant target species, and is determined by an equation analogous to Eq.~(\ref{thetaSumP}).  For illustrative purposes the curves in Figure~\ref{fig:theoIsothermGlobal} are calculated for the case of a single PM/MM pair of probe features addressing the target in question.  Two differences with the local depletion case are immediately apparent.  Firstly, since the depleted targets are shared among more than one probe feature, the asymptotic behaviour of the isotherms at subcritical concentrations differs between different probes addressing the the same target species (i.e.\ $\lim_{X_S\rightarrow\infty} \theta_\PeM > \lim_{X_S\rightarrow\infty} \theta_\MM$).  Secondly, the shape of the hook curve remains unchanged as the probe length varies by the following argument.  Since we use an assumption that $K_S^\PeM/K_S^\MM$ is independent of probe length, one can think of the hook curve as being parameterised by the variable $K_S^\PeM (x_S - p \theta_\Sum)$ where $\theta_\Sum$ has some functional dependence on $x_S$, $p$, $K_S^\PeM$ and the fixed ratio $K_S^\PeM/K_S^\MM$.  Changing the value of $x_S$ then simply effects an identical reparameterisaton in this variable of both $\theta^\PeM$ and $\theta^\MM$.  Individual points will migrate along the path of the hook curve, and at subcritical concentrations the curve will be truncated at the right hand end at different points, but otherwise the shape of the hook curve remains unchanged.    

\subsection{Experiment}

The Suzuki spike-in experiment~\cite{Suzuki07} includes spike-in runs of 150 cDNA target sequences both with and without a complex background.  To keep the analysis simple we analyse only the data set without a complex background.  The data set with complex background provides very similar results with respect to target depletion.  

\begin{figure}
\includegraphics[scale=0.75]{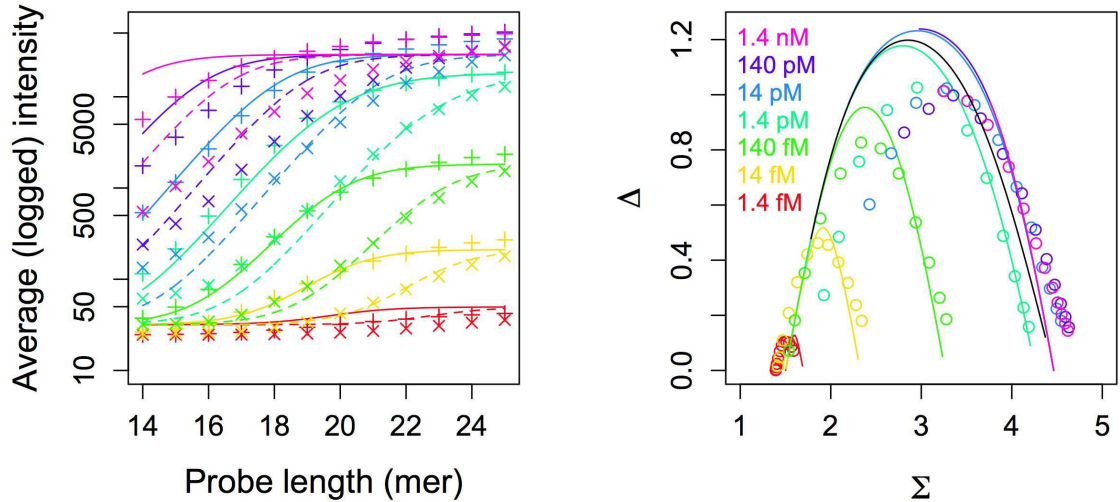}
\centering
\caption{The Suzuki et al. data set without complex background.  Left panel: Fluorescence intensities from PM ($\times$) and MM in the central nucleotide position ($+$) probes obtained by taking average logged intensities of 150 spiked cDNA sequences.  Fits to the model Eq.~\ref{localModel} are shown as a solid curve (PM) and dashed curve (MM).  Right panel: the corresponding hook curves and fits.  The black hook curve corresponds to the fitted critical concentration $x_{\rm crit} = p = 2.26\,$pM. }
\label{fig:SuzukiData}
\end{figure}

In Figure~\ref{fig:SuzukiData} are plotted average logged intensities ($I_{\rm av.log} = 10^{\langle log_{10} I \rangle}$) over three replicates of the 150 sequences for PM and MM probes of varying lengths, the mismatches being in the central position of the probe.  Thus each plotted data point is an average over 450 raw intensities.Some sort of averaging over probe sequences to account for the dependence of binding affinity on individual probe sequences was necessary in order to separate out the dependence on probe length.  This is handled in the implementation of the hook curve for Affymetrix chips by correcting intensities with position- and nucleotide-dependent sensitivity profiles determined from intensity distributions over the whole array~\cite{Binder08b}.  However, this method cannot be used for the Suzuki data set as each array contains a range of probe lengths, making it difficult to define meaningful sensitivity profiles.  The use of average logged intensities rather than averaged intensities was an appropriate and simple solution accounting for the fact that microarray data is generally observed to have multiplicative errors.  Comparison of the resulting hook curves in Figure~\ref{fig:SuzukiData} with the theoretical hook curves in Figures~\ref{fig:theoIsothermLocal} and \ref{fig:theoIsothermGlobal} shows clear evidence for local rather than global depletion.  

Also shown in Figure~\ref{fig:SuzukiData} are fits of a six parameter model to the 168 data points (12 probe lengths $\times$ 7 concentrations $\times$ PM and MM).  The model, based on the theoretical solution  Eq.~(\ref{thetaSoln1}) for local depletion with $X_N = 0$, is defined by 
\begin{equation}
I_{\rm av.log}^P = A + B \theta^P,    \qquad P = \PeM, \MM \label{localModel}
\end{equation}
where $\theta^P$ is the solution to 
\begin{equation}
\theta^P = \frac{\kappa_P e^{\lambda(\ell - 20)}(x - p\theta^P)}{1 + \kappa_P e^{\lambda(\ell - 20)}(x - p\theta^P)},
\end{equation}
$\ell$ is the probe length and $x$ the spike-in concentration.  The parameters $A$ and $B$ account for the optical background intensity and saturation intensity respectively and the effective equilibrium constant in Eq.~(\ref{thetaSoln1}) is modelled by $K_S^P =\kappa_P e^{\lambda(\ell - 20)}$.  The fitted parameter values are listed in Table~\ref{tab:localFit}.  The fitted value of the effective probe concentration $p = 2.26\,$pM is consistent with the observations of Ono et al.~\cite{Ono08}.  

\begin{table}[h]
\caption{\small Parameters fitting the local depletion model Eq.~(\ref{localModel}) to the Suzuki data set.}
\label{tab:localFit}
\begin{center}
\begin{tabular}{lll}
\hline
Optical background intensity                   & $A$ &  31.7\\
Saturation intensity above background & $B$ &  $2.90 \times 10^4$ \\
Equilibrium constant of 20\,mer PM probe & $\kappa_\PeM$  & 0.500\,pM$^{-1}$ \\
Equilibrium constant of 20\,mer MM probe & $\kappa_\MM$  & 0.022\,pM$^{-1}$ \\
Logarithmic length increment of $K_S$ per nucleotide & $\lambda$ & 1.02 \\
Bulk equivalent concentration of probes & $p$ & 2.26\,pM\\
\hline
\end{tabular}
\end{center}
\end{table}

\section{The Affymetrix latin square data set: an example of global depletion}
\label{AffyDataAnalysis}

Affymetrix have produced two well known data sets~\cite{AffyLatinSq} from experiments in which RNA transcripts were spiked in at cyclic permutations of a set of known concentrations together with a complex background of cRNA extracted from human pancreas or human adenocarcinoma cell line and hybridised onto U95a or U133 GeneChips respectively.  In a previous analysis~\cite{Burden08} the U95a data set was shown to fit very well, and the U133 data set moderately well, to a physico-chemical model in which the target concentration was assumed not to be significantly depleted from the supernatant solution by hybridisation to the microarray surface.  This model was the $p = 0$ limit of the models in \ref{sec:PCModel}.  In this section we reanalyse the U133 data set and demonstrate that the global model of target depletion provides a significantly improved fit to this data.  

\subsection{Theory}

The global model of target depletion is relevant to U133 Affymetrix GeneChips as the elements of a PM/MM pair of features are located in neighbouring locations on the microarray surface.  Although each targeted gene is represented by 11 such probe pairs, we ignore depletion from other features within the same probeset as the design of the chip is such that those features are located elsewhere on the chip, and in general will target parts of the gene sequence further removed than the typical target fragment size of about 200 bases.  

The coverage fraction $\theta^P$, $P \in \{\PeM, \MM\}$, of fluorescent dye carrying target molecules bound to the PM or MM feature at completion of the hybridisation step is given by Eq.~(\ref{thetaSolnGlobal}) where $\theta_\Sum = \theta^\PeM + \theta^\MM$ is found by solving Eq.~(\ref{thetaSumP}), and $X_N^P$ and $K_S^P$ are the non-specific binding strength and effective equilibrium constant for specific binding respectively.  The loss of fluorescence intensity due to the post-hybridisation washing step cannot be ignored for Affymetrix GeneChips~\cite{Burden06,Held06,Skvortsov07}, and we introduce into our model specific and non-specific washing factors $w_S^P$ and $w_N^P$ respectively, where $1 > w_S^P > w_N^P > 0$.  The post-washing coverage fraction is then given by Eq.~(\ref{postWashTheta}).  Finally, the observed fluorescence intensity is 
\begin{eqnarray}
I^P  & = & a + b\theta_{\rm after.wash}^P \nonumber \\
        & = & a + b \frac{w_N^P X_N^P + 
                    w_S^P K_S^P (x_S - p \theta_\Sum)}{1 + X_N^P + K_S^P (x_S - p \theta_\Sum)}, 
               \qquad P = \PeM, \MM.        \label{ISolnP}
\end{eqnarray}
where $a$ and $b$ are the physical background and absolute saturation intensities, assumed to be constant across the entire microarray.  

\begin{figure}
\includegraphics[scale=0.75]{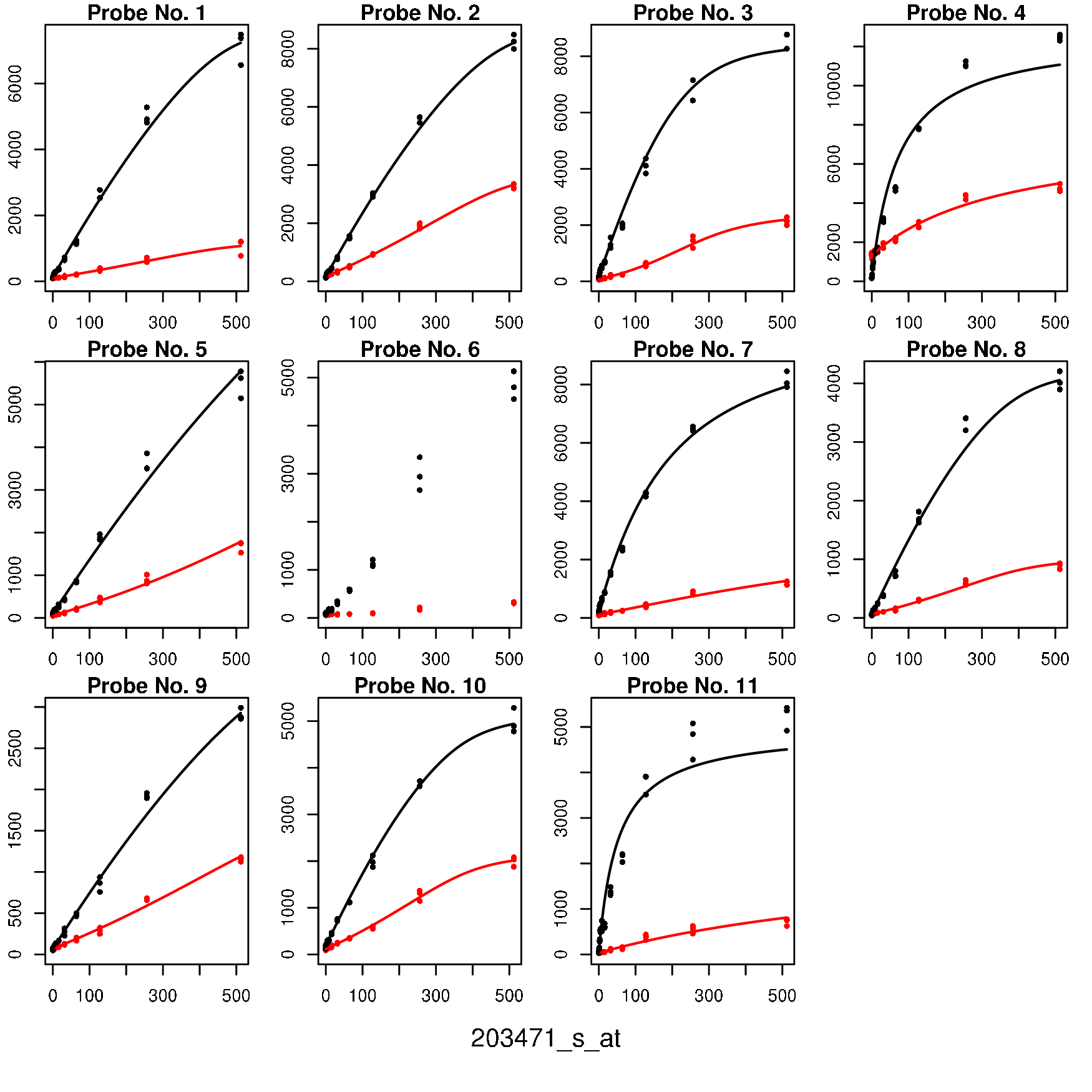}
\centering
\caption{Fits of measured fluorescence intensities in .cel file units against spike-in concentrations in pM from a selected probeset of the spiked transcripts in the Affymetrix latin square U133 experiment to the 7 parameter model defined by Eq.~(\ref{Ifit}).  Note that a flattening of the PM isotherm and an inflection point in the MM isotherm, predicted in Section \ref{sec:ShapeIso} to be a characteristics of target depletion in certain parameter regimes, is clearly visible for most of these probes.}
\label{fig:fit203471s}
\end{figure}

\subsection{Experiment}

For the purposes of comparing fits of the spike-in data to a null-hypothesis model without depletion ($p = 0$) and the one-sided alternate hypothesis with depletion ($p > 0$), we rewrite the model defined by Eqs.~(\ref{ISolnP}) and (\ref{thetaSumP}) in the form 
\begin{equation}
I^P(x) = A^P + B^P \frac{K^P(x - p \theta_\Sum)}{1 + K^P(x - p \theta_\Sum)},  \qquad P = \PeM, \MM \label{Ifit}
\end{equation} 
where $\theta_\Sum(x;K^\PeM,K^\MM,p)$ is the solution in the physically relevant interval $0 \le \theta_\Sum \le 2$ to 
\begin{equation}
\theta_\Sum = \sum_{P = \PeM, \MM} 
\frac{K^P (x - p \theta_\Sum)}{1 + K^P (x - p \theta_\Sum)}.  \label{thetaSumFit}
\end{equation}
Here we have suppressed the subscript $S$ on the PM-specific spike-in concentration $x_S$ and introduced the parameterisation 
 \begin{eqnarray}
A^P & = & a + b w_N^P \frac{X_N^P}{1 + X_N^P}, \\
B^P & = & b \left( w_S^P -  w_N^P \frac{X_N^P}{1 + X_N^P} \right), \\
K^P & = & \frac{K_S^P}{1 + X_N^P}. \qquad P = \PeM, \MM.
\end{eqnarray}

Equations (\ref{Ifit}) and (\ref{thetaSumFit}) define a 7 parameter model ($A^P$, $B^P$, $K^P$, $p$) to which intensity data  from a PM/MM pair of features for a range of spike-in concentrations $x$ can be fitted.  The $p = 0$ case, corresponding to no significant target depletion, defines a 6 parameter model which was previously fitted to the U95a data in ref.~\cite{Burden04} and to both the U95a and U133 data in ref.~\cite{Burden06}.  Below we use standard statistical methods to distinguish between a null hypothesis, $p = 0$, and alternate hypothesis $p > 0$.  

Fluorescence intensities for each of 11 probe pairs from each of 38 spike-in transcripts of the U133 latin square experiment were fitted assuming the data to be Gamma distributed with mean given by the model of Eq.~(\ref{Ifit}).  The assumption of Gamma distributed data was used in previous analyses~\cite{Burden04} to accommodate a constant coefficient of variation as expected for data with multiplicative errors, and is easily implemented using the function {\tt glm()} from the statistical computing environment R~\cite{RCranWebPage}.  Fits of the model to the data of one of the spiked transcripts are plotted in Figure~\ref{fig:fit203471s}, and analogous plots for all 38 spiked transcripts are available in the supplementary material or at the web site of one of the authors~\cite{BurdenSpikeInPage}.  

\begin{figure}
\includegraphics[scale=0.75]{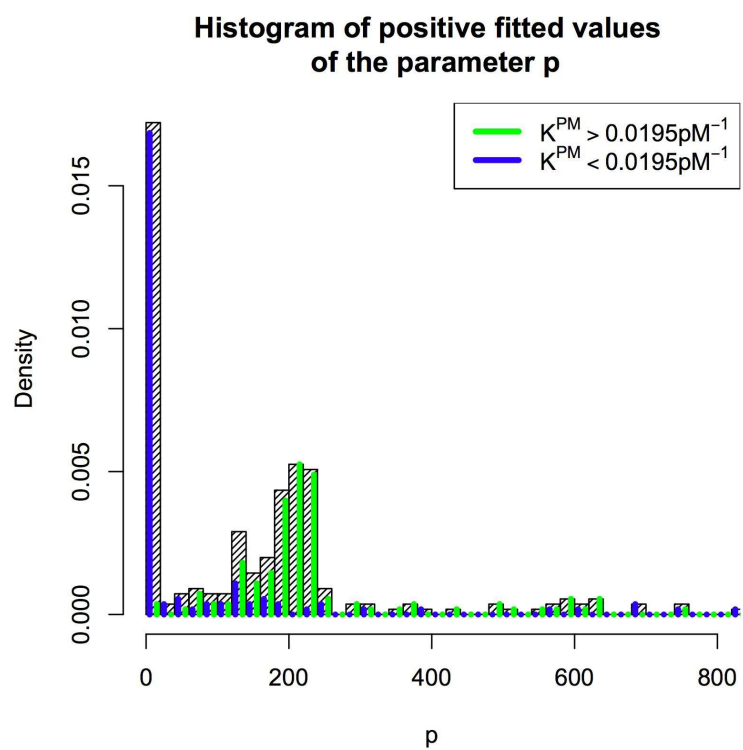}
\centering
\caption{Histogram of the fitted value of the parameter $p$ (pM) for the all of the 276 fits with physically meaningful parameter values (hatched bars).  Also shown are histograms of the subsets corresponding to high and low values of the parameter $K^\PeM$.  The low-$K^\PeM$ probes correspond to data lying to the left of the vertical dotted line in Figure~\ref{fig:effectiveX} and are an  approximation to the set of probes for which depletion data can also be fitted to a no-depletion model with an effective equilibrium constant given by Eq.~(\ref{underestimateK}).}
\label{fig:fittedP}
\end{figure}

Of the 418 probe pairs in the data set, 276 (or 66.0\%) were successfully fitted to physically relevant values of the effective probe concentration restricted to the range of concentratons $p \ge 0$ with physical values for the remaining parameters, i.e., $A_\PeM$, $B_ \PeM $, $K_ \PeM $, $A_\MM$, $B_\MM$ and $K_\MM$ all $> 0$.  This should be compared with fits to the $p = 0$ model without depletion in ref.~\cite{Burden08}, for which only 37.5\% of probes were successfully fitted to PM/MM probe pairs.  A histogram of the fitted values of the effective probe concentration parameter $p$ is shown in Figure~\ref{fig:fittedP}.  

\begin{figure}
\includegraphics[scale=0.75]{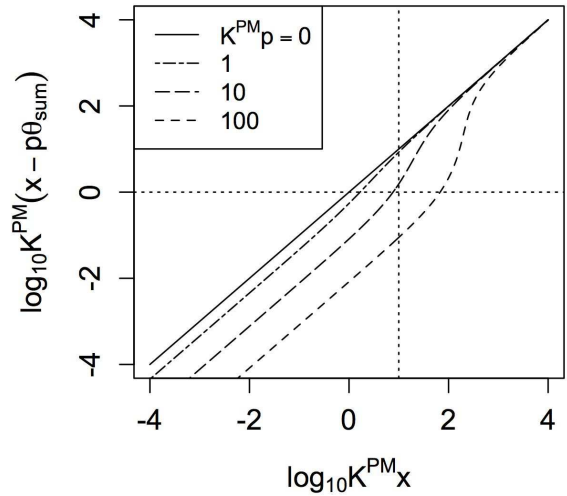}
\centering
\caption{The relationship between the effective binding strength $K^{\PeM}x_{\rm eff}$ and true binding strength $K^{\PeM}x$ for a range of effective probe concentrations.  The curves are calculated with the help of Eq.~(\ref{thetaSumFit}), assuming $K^\PeM/K^\MM = 5$, though in practice the shape of the curves is not very sensitive to this ratio.  The horizontal dotted line is the upper limit of binding strengths for which depletion data can also be fitted to a no-depletion model with an effective equilibrium constant given by Eq.~(\ref{underestimateK}).  The vertical dotted line is the right hand limit of binding strengths determining the set of low-$K^\PeM$ probes in Figure~\ref{fig:fittedP}.}
\label{fig:effectiveX}
\end{figure}

Immediately noticeable is that the distribution is bimodal: a number of fits are simply the `no depletion' solution at $p = 0$, while most of the the remaining cases cluster around $p=200\,{\rm pM}$.  To understand this, note that Eq.~(\ref{Ifit}) makes clear that the effect of depletion is to reduce the true target concentration $x$ to an effective concentration 
\begin{equation}
x_{\rm eff} = x - p \theta_\Sum,   \label{xEff}
\end{equation}
where $\theta_\Sum$ is the sum of the PM and MM hybridisation fractions due to specific binding only, and is obtained by solving Eq.~(\ref{thetaSumFit}).  In Figure~\ref{fig:effectiveX} is plotted the corresponding effective binding strength $K^{\PeM}x_{\rm eff}$ against the true binding strength $K^{\PeM}x$.  One sees that, below a certain binding strength indicated by the horizontal dotted line $K^{\PeM}x_{\rm eff} = 1$, the true concentration is reduced by a factor which is approximately constant over a range of $x$.   In fact, from Eq.~(\ref{thetaSumFit}) we have that, for $K^\PeM x_{\rm eff} \ll1$, i.e. the linear, low-concentration part of the isotherm, $\theta_\Sum \approx (K^\PeM + K^\MM)x_{\rm eff}$, from which it follows using Eq.~(\ref{xEff}) that $x_{\rm eff} \approx x/[1 + (K^\PeM + K^\MM)p]$.  It follows that any probe whose data points lie within this range will be fitted equally well by a hyperbolic, no-depletion, isotherm $I^P = A^P + B^P K_{\rm eff}^P x/(1 + K_{\rm eff}^P x)$, with an underestimated equilibrium constant: 
\begin{equation}
K_{\rm eff}^P = \frac{K^P}{1 + (K^\PeM + K^\MM)p}, \qquad P=\PeM, \MM.  \label{underestimateK}
\end{equation}

In Figure~\ref{fig:fittedP} we have partitioned the fitted values of $p$ into those matching with high and low fitted values of the equilibrium constant, $K^\PeM \gtrless \left(\frac{10}{512} \approx 0.0195\right){\rm pM}^{-1}$ respectively.  The cutoff is chosen as a simple way to separate out an approximate set of probe pairs satisfying the conditions leading to the result of Eq.~(\ref{underestimateK}): Recall that the spike-in concentrations in the U133 experiment are bounded above by 512\,pM, so for the low-$K^\PeM$ probes $\log_{10} K^\PeM x < 1$.  That is, the fitted isotherms of these probes are determined solely from data lying to the left of the vertical dotted line in Figure~\ref{fig:effectiveX} for which the curves relating  $\log x$ to $\log x_{\rm eff}$ are approximately straight.  Returning to Figure~\ref{fig:fittedP}, one observes that the high-equilibrium-constant isotherms, $K^\PeM > 0.195\, {\rm pM}^{-1}$, fit predominantly to the depletion model with $p$ consistently around $p=200\,{\rm pM}$, and the low-equilibrium-constant isotherms fit predominantly to the no-depletion model with, we infer, the fitted parameter $K^P$ underestimated according to Eq.~(\ref{underestimateK}).  A rough estimate of the lower limit of the underestimation factor, assuming $K^\MM \ll K^\PeM$,  is $(1 + 0.0195)^{-1} \times 200 \approx 0.2$.  

\begin{figure}
\includegraphics[scale=0.75]{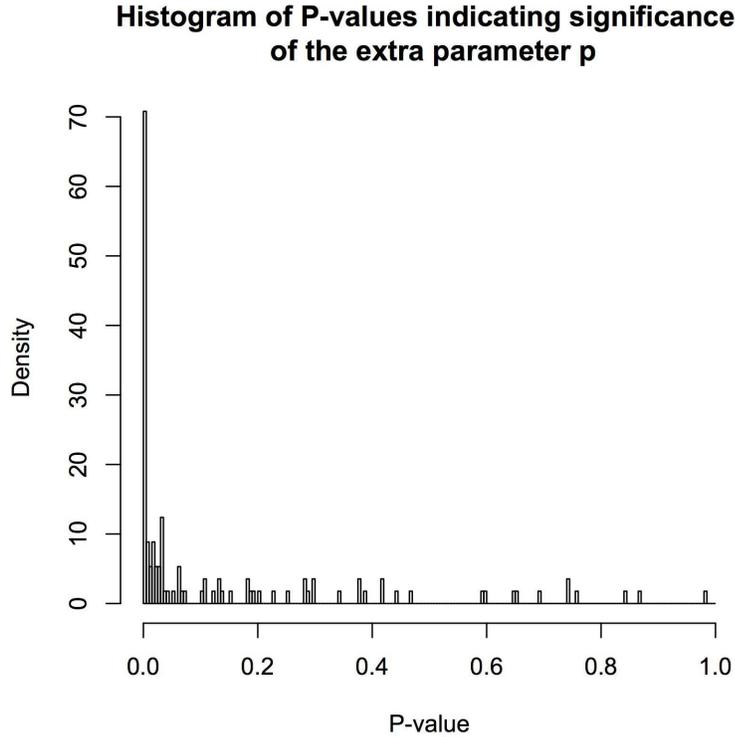}
\centering
\caption{Histogram of the fitted P-values under the null hypothesis of no target depletion ($p = 0$) tested against the alternate hypothesis of global target depletion ($p > 0$) for each of those probe pairs which admit physically meaningful fits to both models.  Just over 60\% of cases fall within the 5\% confidence interval (P-values $< 0.05$), favouring the alternative hypothesis.}
\label{fig:significance}
\end{figure}

For the subset of probe pairs which admit physically meaningful fits to both the alternate hypothesis model with depletion, and to the null hypothesis model without depletion, and for which the fitted value of $p$ is strictly positive, we calculated one sided P-values under the null hypothesis assumption using the analysis appropriate to generalised models~\cite{McCullagh89} described in detail in ref.~\cite{Burden04}.   The histogram of these P-values, Figure~\ref{fig:significance}, shows that they are heavily bunched to the left: Depletion is confirmed at the 5\% confidence level for just over 60\% of those cases for which the comparison could be made.  

\subsection{Shape of the isotherms}
\label{sec:ShapeIso}

\begin{figure}
   \centering
   \includegraphics[height=8.0cm]{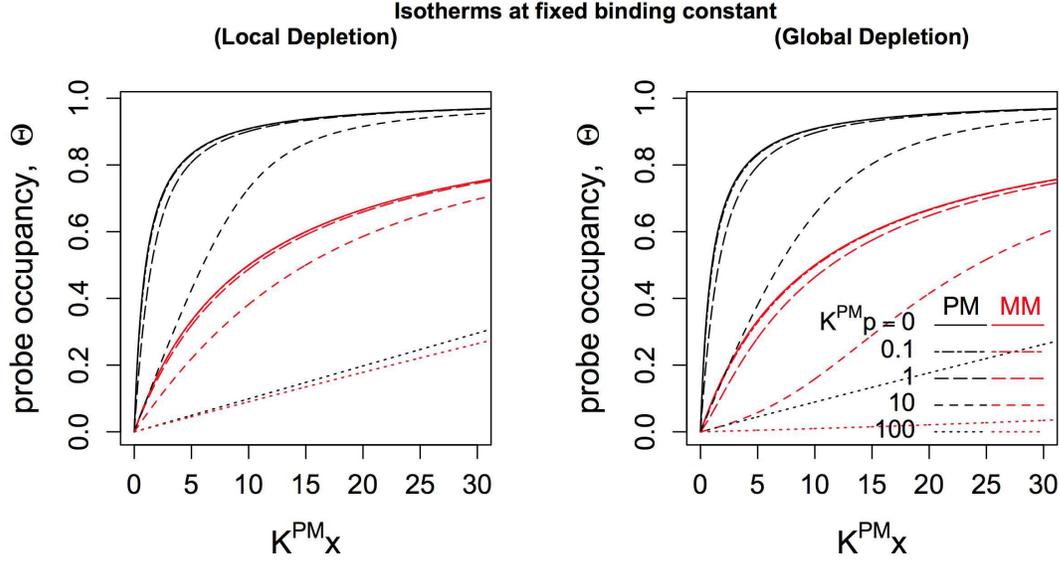} 
   \caption{Theoretical isotherms for PM and MM probes derived from local (left) and global (right) depletion models.}  The isotherms are scaled to dimensionless units $K^\PeM x$, $\Theta^P = (I^P - A^P)/B^P$ for $s = 10$ and various values of the dimensionless depletion factor $K^\PeM p$. As explained in the text, in the global PM/MM model, the MM isotherms have an inflection point for $K^\PeM p > (s - 1)^{-1}$, whereas the PM isotherms do not have an inflection point for any value of this parameter.  Isotherms from the local model have no inflection point for any parameter values.  Note that these isotherms are plotted at fixed values of binding constant $K^\PeM$, whereas the isotherms in Figures~\ref{fig:theoIsothermLocal} and \ref{fig:theoIsothermGlobal} are plotted at fixed values of specific target concentration $x$, and consequently have different asymptotic properties as $K^\PeM x \rightarrow \infty$.  
   \label{fig:isotherms_scaled}
\end{figure}

It is interesting to examine the shape of the isotherm fits in the global PM/MM depletion model to see how they they differ from the well known hyperbolic Langmuir form of the model without depletion, and from the isotherms of the local depletion model.  It is convenient to define dimensionless quantities 
\begin{equation}
\Theta^P = \frac{I^P(x) - A^P}{B^P}, \qquad s = \frac{K^\PeM}{K^\MM}.
\end{equation}
On physical grounds we expect $s>1$, which is observed in general in fits of spike-in data to models with and without depletion.  Eqs.~(\ref{Ifit}) and (\ref{thetaSumFit}) become 
\begin{equation}
\Theta^\PeM = \frac{K^\PeM(x - p\theta_\Sum)}{1 + K^\PeM(x - p\theta_\Sum)}, \qquad
\Theta^\MM   = \frac{K^\PeM(x - p\theta_\Sum)}{s + K^\PeM(x - p\theta_\Sum)},  \label{ThetaP}
\end{equation}
with $\theta_\Sum$ the solution to 
\begin{equation}
\theta_\Sum = \frac{K^\PeM(x - p\theta_\Sum)}{1 + K^\PeM(x - p\theta_\Sum)} 
                       + \frac{K^\PeM(x - p\theta_\Sum)}{s + K^\PeM(x - p\theta_\Sum)}.  \label{thetaSumDimsionless}
\end{equation}
Plots of $\Theta^P$ as a function of the dimensionless $K^\PeM x$ for the realistic value $s=10$ and a range of 
values of the dimensionless depletion parameter $K^\PeM p$ are shown in the right panel of Fig.~\ref{fig:isotherms_scaled}.  Also shown for comparison (left panel) are the equivalent isotherms from the local depletion model, for which $\theta_\Sum$ in Eq.~(\ref{ThetaP}) is replaced by $\Theta^\PeM$ or $\Theta^\MM$ respectively.  The effect of depletion is to depress the response function at small specific target concentrations, as the available effective specific target concentration is effectively decreased.  For the case of the PM/MM global depletion model, we show in \ref{sec:IsoShape} that for $K^\PeM p > (s - 1)^{-1}$, and provided $s > 1$, the MM response curve acquires an inflection point, while the PM curve flattens without forming an inflection point.  Physically, the effect of depletion on the MM response is more pronounced as the PM probes more strongly deplete the available target in solution.  This behaviour is clearly evident in fits to the U133 spike-in data (see Figure~\ref{fig:fit203471s} and the supplementary material).  A straightforward calculation shows that isotherms from the local depletion model, on the other hand, do not have an inflection point for either PM or MM probes for any parameter values.  

\subsection{Shape of the hook curve}
\label{sec:ShapeHook}

\begin{figure}
   \centering
   \includegraphics[height=8.0cm]{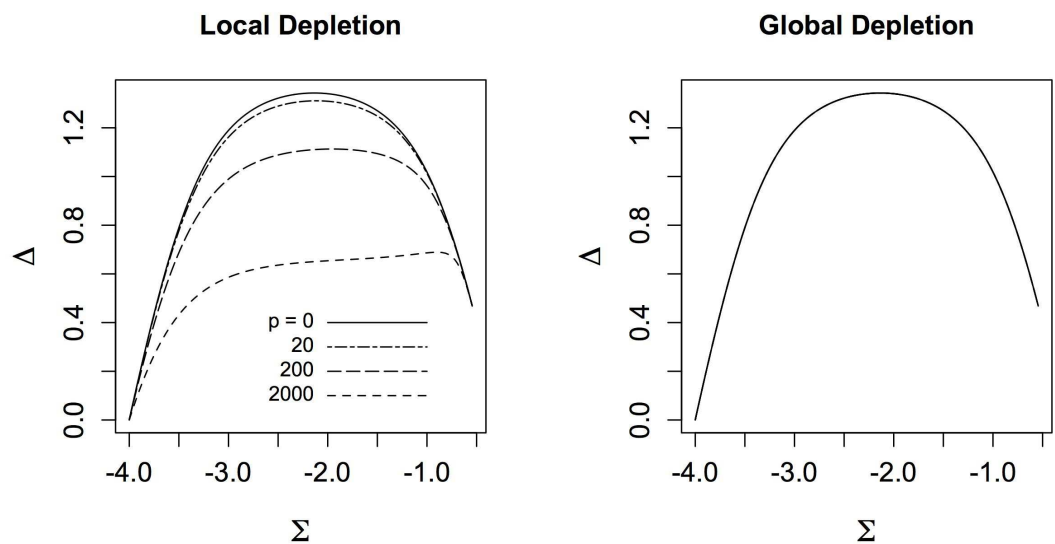} 
   \caption{Theoretical hook curves determined from isotherms of a PM/MM pair assuming local depletion (left) and global depletion (right) for a range of the effective probe concentration parameter $p$ (pM).  The following parameter values, typical of fits to the U133 Latin Square spike-in data set, were used: $X_N^\PeM = X_N^\MM = 10^{-3}$, $K_S^\PeM = 5\times 10^{-3}$ pM$^{-1}$, $K_S^\MM = 5\times 10^{-4}$ pM$^{-1}$, and washing survival fractions $w_N^\PeM = w_N^\MM = 0.1$, $w_S^\PeM = 0.5$, $w_S^\MM = 0.2$.  Note that that for global depletion the shape of the hook curve is independent of $p$.}
   \label{fig:theoreticalSpikeHook}
\end{figure}

Theoretical hook curves assuming either a local or global depletion model and parameter values typical of fits to the U133 Latin Square data set and a range of the probe density parameter $p$ are shown in Fig.~\ref{fig:theoreticalSpikeHook}.  For these curves the trajectory is that of the pair $(\Sigma,\Delta)$ defined by Eq.~(\ref{SigmaDeltaDef}) traced out as the specific binding strength $X_S^\PeM = K_S^\PeM x_S$ varies over a range of specific spike-in concentrations $x_S$ at fixed values of all other parameters in the model.  For the case of global depletion the hook coordinates are calculated from the post-washing coverage fractions Eq.~(\ref{postWashTheta}) with $\theta_\Sum$ given by Eq.~(\ref{thetaSumP}).  For the case of local depletion $\theta_\Sum$ is replaced by $\theta_S^\PeM$ or $\theta_S^\MM$ respectively defined by Eq.~(\ref{thetaSfromQuad}).  

One sees that the effect of local depletion is to flatten the peak and introduce an asymmetry in the hook curve.  The flattening is caused by a decrease in the difference between the PM and MM responses as more specific target is extracted from solution in the vicinity of the PM probe feature.  Global depletion, on the other hand, has no effect on the shape or end points of the hook curve as it effects an identical reparameterisation $x_S \rightarrow x_S - p\theta_\Sum$ in the formulae for both $\theta^\PeM$ and $\theta^\MM$.  However, as $p$ is increased, internal points corresponding to a given probe-pair value of the binding strength migrate progressively to the left along the curve, reflecting a decrease in both the PM and MM fluorescence intensities.   

\begin{figure}
   \centering
   \includegraphics[height=8.0cm]{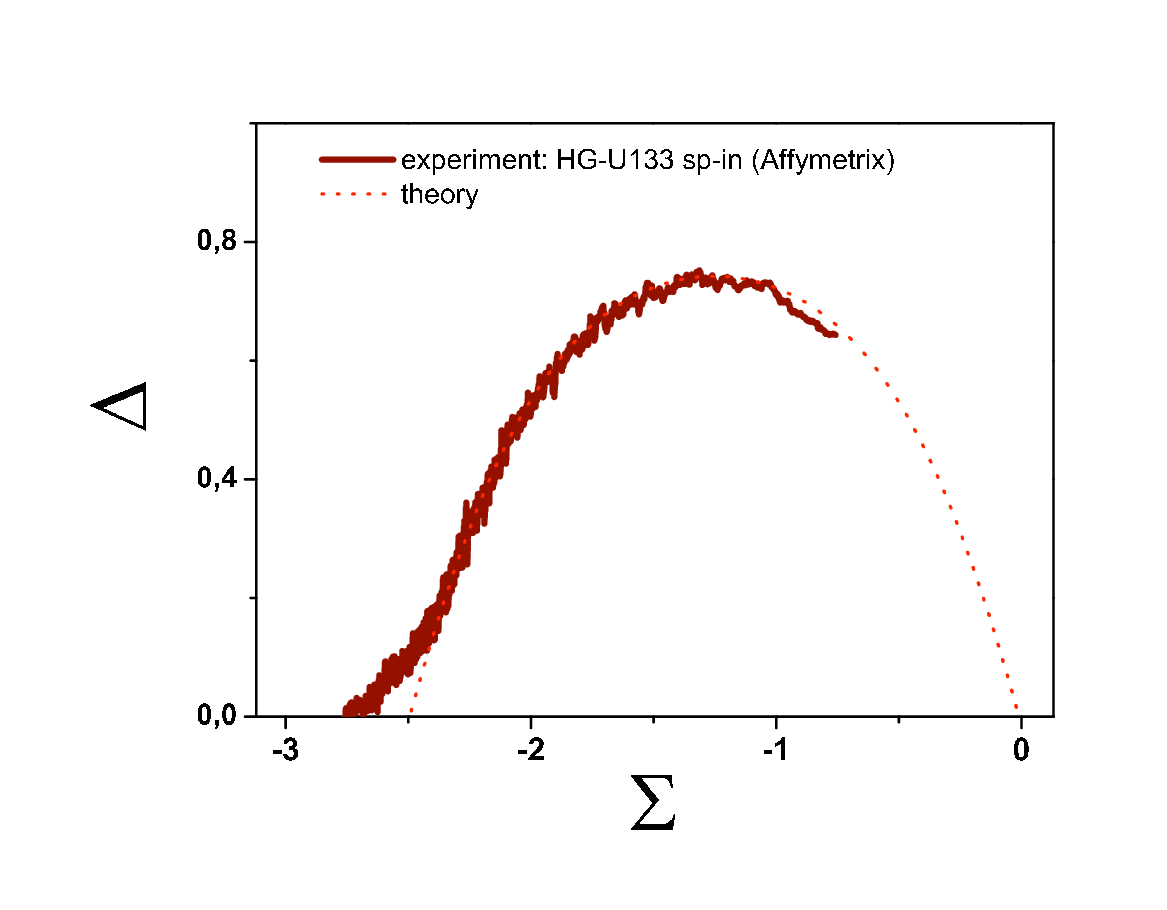} 
   \caption{Experimental hook curve of one array of the U133 Latin Square data set and a fit using assuming the hyperbolic Langmuir response function without depletion. Note the symmetric shape of the experimental hook curve, compatible with global depletion. The deviation between the experimental and theoretic curve at small $\Sigma$ is caused by non-specific hybridisation not discussed here (see~\cite{Binder08a, Binder08b}).}
   \label{fig:U133ExperimentalHook}
\end{figure}

A typical hook curve from one of the arrays of the U133 Latin Square data set using the algorithm at ref.~\cite{HookProgram} is shown in Fig.~\ref{fig:U133ExperimentalHook}.  This algorithm includes a moving average over $\sim 100$ probesets and correction of raw intensities for probe binding affinities using position- and nucleotide-dependent sensitivity profiles~\cite{Binder08b}.  Hook curves have been similarly evaluated for a number of experimental datasets relating to Affymetrix GeneChips in ref.~\cite{Binder08a}, including the Latin Square spike-in experiments, with the result that no evidence for an asymmetric hook curve has yet been observed.  We conclude, within this particular set of data sets corresponding to chip designs with neighbouring PM/MM pairs, that target depletion, if significant, fits the global model rather than the local model.  

\subsection{Correction of expression estimates}
\label{sec:Correction}

Estimates of expression levels using algorithms such as the hook method~\cite{Binder08a} and the inverse Langmuir method~\cite{Mulders09} have to date ignored target depletion and therefore been based on the assumption of a hyperbolic Langmuir isotherm.  In the previous section we have seen that this is equivalent to underestimating the true specific target concentration $x_S$ by a shift $x_S \rightarrow x_{\rm eff} = x_S - p\theta_\Sum$, where $\theta_\Sum$ is the sum of the PM and MM hybridisation fractions due to specific binding only.  $\theta_\Sum$ can be calculated from the observed total coverage fractions $\theta^\PeM$ and $\theta^\MM$, which include both specific and non-specific binding, as follows:  

From Eq.~(\ref{thetaSumP}),
\begin{equation}  
\theta_\Sum = \frac{K_S^\PeM x_{\rm eff}}{1 + X_N + K_S^\PeM x_{\rm eff}} + \frac{K_S^\MM x_{\rm eff}}{1 + X_N + K_S^\MM x_{\rm eff}},   \label{thSum}
\end{equation}
where the nonspecific strength $X_N = X_N^\PeM \approx X_N^\MM$ is assumed to be common for all probes on the microarray after correction for binding affinities via sensitivity profiles.  $X_N$ can be measured from the width of the hook curve and is typically of order $10^{-3}$.  From Eq.~(\ref{thetaSolnP}), 
\begin{equation}
\theta^P = \frac{X_N + K_S^P x_{\rm eff}}{1 + X_N + K_S^P x_{\rm eff}}, \qquad P = \PeM, \MM,  
\end{equation}
which rearranges to give $K_S^P x_{\rm eff} = \theta^P/(1 - \theta^P) - X_N$.  Substituting back into Eq.~(\ref{thSum}) then gives $\theta_\Sum = (1 + X_N)(\theta^\PeM + \theta^\MM) - 2X_N$.  

Thus, the true specific target concentration is given in terms of the effective, depleted target concentration by
\begin{equation}
x_S = x_{\rm eff} + p\left[(1 + X_N)(\theta^\PeM + \theta^\MM) - 2X_N\right].  \label{xSCorrection}
\end{equation}
In principle, this formula gives the correction for target depletion over the entire range of target concentrations, including an interpolation between the two regimes illustrated in Figure~\ref{fig:effectiveX}.  Note that $x_{\rm eff}$, $X_N$ and the coverages $\theta_P$ can be estimated by established methods such as hook curve or inverse Langmuir method.  Eq.~(\ref{xSCorrection}) then requires knowledge of  the probe concentration $p$, which, for example, is expected to depend on the chip type.  Its estimation requires further efforts which will be the subject  of future investigations

\section{Critical evaluation of an alternate hybridisation model}
\label{sec:CritEval}

Recently an alternate competitive hybridisation model incorporating target depletion by Li et al.~\cite{Li08} has appeared in the literature.  This model is applied to the Affymetrix U133 data set and is purported to be capable of predicting signal intensities of individual probes and of achieving quantification of absolute target concentrations from microarray fluorescence intensity data.  Here we point out a number of errors in the basic assumptions of Li et al.'s model and argue that it does not represent any advance over previously existing hybridisation models.  

Of particular interest to Li et al.\ is the asymptotic behaviour of fluorescence intensities for individual probes in the limit of saturation concentrations of specific target.  Standard reaction kinetic models applied to the hybridisation step of the Affymetrix protocol implies that in the high specific target concentration limit, all probes should saturate at the same observed fluorescence intensity, regardless of the nucleotide probe sequence or resulting probe-target binding free energy.  For either of the models in \ref{sec:PCModel}, for instance, we have $\lim_{x_S \rightarrow \infty}\theta =1$, where the limit is taken with other variables being held constant.  This is at variance with observations from spike-in experiments, for which the PM element of a PM/MM almost invariably saturates at a higher intensity than its MM partner.  

An acceptable explanation, which has been demonstrated to fit well the saturation behaviour to both the U95a and U133 Affymetrix spike-in experiments~\cite{Burden06,Held06}, is to explain the differing asymptotes via the post-hybridisation washing step, which not only removes unbound targets, but also dissociates both specific and non-specific bound targets (see Eq.~(\ref{postWashTheta})).  For reasons which are not clear, but which appear to be based on a misinterpretation of Skvortsov et al.'s experimental results~\cite{Skvortsov07}, Li et al.\ reject the washing hypothesis.  Instead, they proceed to develop their own thermodynamic model, which is not consistent with accepted principles of physical chemistry, but which nevertheless predicts response functions with binding free energy dependent asymptotes resulting from the hybridisation step alone.  In their model, the washing step is assumed to have little effect on specific targets bound to probes.  

In \ref{sec:LiCritique} we explain in detail a fundamental error in their application of the law of mass action to hybridisation at the microarray surface, and show that when the error is corrected, their model essentially agrees with existing treatments inspired by Langmuir adsorption theory, together with the depletion extension of Ono et al.~\cite{Ono08}.  We also note that their derived formula for the coverage fraction of specific targets is demonstrably wrong in that it disagrees with the results of the Affymetrix latin square spike-in experiments without  complex background.  Lastly, the algorithm proposed by Li et al.\ for inferring absolute specific target concentrations requires subtraction of the intensity at zero spike-in concentration as a way of dealing with non-specific hybridisation (see Eqs.~(22) and (27) of ref~\cite{Li08}).   This value is of course unknown in any biomedical application of microarrays, and it is the problem of calibrating a correction for non-specific hybridisation which is the subject of much current activity in physico-chemical modelling of microarrays (see refs.~\cite{Kroll08, Binder09a} for instance).  That Li et al.\ are able to produce estimates of spike-in concentrations at the higher end of the scale ($> 1$\,pM) by cross validation from a crude 4 parameter formula based on incorrect physical assumptions is not surprising and is not an improvement on any existing expression measure.  

\section{Conclusions and outlook}
\label{sec:Conclusions}

The physico-chemical models of equilibrium microarray hybridisation described here involve microarray probes, specific and non-specific targets and their interactions on the chip surface.   As well as probe-target hybridisation, bulk hybridisation and probe and target folding, the important innovation is a careful consideration of depletion of target molecules from the supernatant solution by hybridisation of specific targets.  Consideration of target depletion is important when the target concentration is comparable with or less than the effective probe molecule concentration, which we determine to be of the order of 200\,pM for HG133 generation Affymetrix GeneChips.  If the sensitivity of microarrays is to be pushed to lower specific target concentrations, a proper understanding of and appropriate correction for this phenomenon is important.  

Two possible scenarios are considered, local and global depletion.  In the first scenario, studied by Ono et al.~\cite{Ono08}, depletion by hybridisation to a given probe feature only affects that particular feature. This scenario is relevant when probe features addressing the same target species are physically separated on the microarray, and the rate of diffusion or convection over the distance between features is small compared with the rate of hybridisation.  The second scenario, global depletion, has not been considered previously.  In this scenario some or all of the features addressing a given target species are effected.  This is relevant, for instance, for chip designs which include mismatch features located in close proximity on the microarray surface to their perfect match partners.  

We analysed data obtained in two experimental situations:  firstly, the intensity response of PM and MM probes of varying probe length at fixed target concentration (the Suzuki et al.\ data set), and secondly, the intensity response of PM and MM probes of fixed length at varying target concentration (the Affymetrix latin square data set).  The PM/MM design of the chips allows for a combined analysis of both probe types via the ``hook plot'', the shape of which gives a clear discrimination between local and global depletion.  

We have confirmed conclusively using the hook curve analysis that the spike-in data set of Suzuki et al.~\cite{Suzuki07} is an example of local and not global depletion.  A six parameter fit of the local depletion model verifies the earlier analysis of Ono et al.~\cite{Ono08}.  The hook curve analysis has proved particularly useful for this type of analysis because of the marked qualitative difference in the behaviour of these plots between the two possible scenarios.  

Previous attempts to fit a hyperbolic Langmuir isotherm model to the second data set, the Affymetrix U133 latin square spike-in, had only met with partial success~\cite{Burden08}.   In our current reanalysis of this data set we have had markedly improved success using the global model of target depletion, which we believe is relevant because of close proximity of partner PM and MM probe features.  The global depletion model provides a significantly improved fit to a large portion of these data, namely that portion for which the effective equilibrium constant of the hybridisation reaction is above a certain threshold value.  Importantly, we have demonstrated that if the effective equilibrium constant $K$ is below the inverse of the range of concentrations of a spike-in experiment, the ability to detect target depletion through response curve fits is masked and the data may mistakenly be fitted to the linear part of a non-depleted hyperbolic Langmuir isotherm with an underestimated equilibrium constant given by Eq.~(\ref{underestimateK}).  

For the Affymetrix spike-in data our depletion model is also able to explain certain qualitatively observed phenomena related to the shape of the isotherms.  The MM response function typically has an inflection point at low concentrations which may serve as a signal for global depletion in spike-in experiments, whereas the PM response function is typically flattened but does not form any such inflection point.     Another characteristic of global depletion is the shape of the hook curve which continues to be symmetric as the effective concentration of free specific targets is reduced by hybridisation.  Local depletion, on the other hand, is predicted to entail an antisymmetric hook curve.  

In the final section we have given a critique pointing out a number of serious errors in a competing physico-chemical model dealing with target depletion in microarray hybridisation experiments by Li et al.  After correction of these errors one gets a solution which, in the limit of no depletion, is the well established and accepted Langmuir model.  With depletion included it is a simplified version of our local depletion model or the model of Ono et al.~\cite{Ono08}.  

The observations made herein, particularly those for the Affymetrix U133 data set, have consequences for existing physico-chemistry-based algorithms and methods for microarray calibration.  By calibration we mean obtaining estimates of transcript abundance, ideally as an absolute concentration or, at the very least, relative measures which are related linearly to transcript abundance.  It must include not only systematic correction for the effects such as non-specific background, saturation and sequence-specific binding affinities of probes~\cite{Binder09b}, but also, as we have shown, depletion of targets from the supernatant solution.  

Physico-chemical calibration algorithms rely directly or indirectly on obtaining estimates of the effective equilibrium constant $K$ from probe sequences via position dependent affinities~\cite{Binder05,Binder06b} or via free binding energies $\Delta G$ calculated from nearest neighbour stacking models~\cite{Heim06, Heim06b}.  To date they have assumed a hyperbolic Langmuir isotherm and involve fits to spike-in data sets including the Affymetrix HGU133 data set.  We have shown here that estimates of $K$ from this data set are compromised in a predictable way by target depletion if a hyperbolic isotherm is assumed.  It is consequently not surprising that attempts to find a clear and unambiguous relationship between $K$ obtained in this way and $\Delta G$ have met with limited success (see Section 5.2 and 5.3 of~\cite{Burden08}).   Clearly more work has to be done in correcting this aspect of calibration algorithms to take into account target depletion.  Finally, irrespective of whether calibration algorithms rely on inverting a theoretical isotherm~\cite{Mulders09} or first extracting an effective binding strength $X_{\rm eff} = Kx_{\rm eff}$ from, say, the hook curve~\cite{Binder09b}, a solution must be found to the problem of extracting the true target concentration $x$ from the microarray-depleted concentration $x_{\rm eff}$.  In Section~\ref{sec:Correction} we show that the information required to do this is, in principle, inherent in the measured fluorescence intensities via Eq.~(\ref{xSCorrection}).  A practical implementation of this will be the subject of future work.  

\section*{Acknowledgements}

This work was supported in by an Australian Research Council Discovery Project Grant (DP0987298), an Australian Academy of Science Scientific Visits to Europe Grant and by Deutsche Forschungsgemeinschaft (BIZ6-06).

\appendix

\section{Physico-chemical model}
\label{sec:PCModel}

In the physico-chemical model presented below the equilibrium coverage fraction $\theta$ ($0 \le \theta \le 1$) of fluorescent dye carrying target molecules bound to a given probe feature at the microarray surface at the end of the hybridisation step is calculated assuming standard equilibrium physical chemistry.  The model differs from previous models considered by the current authors~\cite{Binder06a, Burden08} in that the Ono model~\cite{Ono08} of target depletion from the supernatant solution by hybridisation to the array is included.  Two regimes are considered: 

The first of these, local depletion, in which depletion by a given probe feature only affects that particular probe, is a slight variant of the ``finite hybridisation model'' including competitive specific and non-specific hybridisation presented by Ono et al.~\cite{Ono08}.  It differs from the Ono model in that all chemical reactions, viz.\ folding, bulk hybridisation and surface hybridisation, are integrated {\em ab initio}, leading to slightly different formulae for the final coverage fraction.  A detailed derivation of local depletion is included here for completeness and to establish the notation and a framework for the second regime, global depletion, in which all features responding to a given target species are affected.  

\subsection{Local Depletion} 
\label{sec:ALocalDepletion}

In this case there is assumed to be no interaction between different probe features.  The set of chemical species considered is set out in Table~\ref{tab:species}.  For a given probe feature, the input parameters to the model are (1) the total specific target concentration,
\begin{equation}
x_S = [S] + [S'] + [P.S] + [S.N] + [S.S],    \label{xS}
\end{equation}
(2) an effective total non-specific target concentration, assumed to be common to all probe features: 
\begin{equation}
x_N = [N] + [N'] + [P.N] + [S.N] + [N.N],    \label{xN}
\end{equation}
(3) an effective probe concentration for the feature:
\begin{equation}
p = [P] + [P'] + [P.S] + [P.N],                        \label{pTotal}
\end{equation}
and (4) a set of equilibrium constants $K_r$, where $r \in \{\Sfold, \Nfold, \ldots\}$, for the reactions (\ref{K_Sfold}) to (\ref{K_PN}) set out below. Following the usual convention square brackets indicate the molar concentration of a chemical species.  

\begin{table}[h]
\caption{\small Chemical species present in the model.}
\label{tab:species}
\begin{center}
\begin{tabular}{lcc}
\hline
	& unfolded & folded \\
specific target in solution 				& $S$ & $S'$ \\
non-spec.\ effective target in solution 	& $N$ & $N'$ \\
probe at surface (not bound to target)	 & $P$ & $P'$ \\
\hline
duplexes in solution					&  \multicolumn{2}{c}{$S.S$, $S.N$, $N.N$}\\
duplexes at microarray surface			&  \multicolumn{2}{c}{$P.S$, $P.N$}\\
\hline
\end{tabular}
\end{center}
\end{table}

Our aim is to determine the total coverage fraction 
\begin{equation}
\theta = \theta_S + \theta_N = \frac{[P.S]}{p} + \frac{[P.N]}{p}.   \label{thetaDef}
\end{equation}
of both specific and non-specfic duplexes resulting from the following chemical reactions: \\
\noindent Folding
\begin{eqnarray}
S \rightleftharpoons S' : & \qquad [S'] = K_{\Sfold} [S].       \label{K_Sfold} \\
N \rightleftharpoons N' : & \qquad [N'] = K_{\Nfold} [N].       \label{K_Nfold}  \\
P \rightleftharpoons P' : & \qquad [P'] = K_{\Pfold} [P].       \label{K_Pfold} 
\end{eqnarray}

\noindent Bulk hybridisation
\begin{eqnarray}
S + N \rightleftharpoons S.N : & \qquad [S.N] = K_{\SN} [S][N].       \label{K_SN} \\
S + S \rightleftharpoons S.S : & \qquad [S.S] = K_{\SeS} [S]^2.       \label{K_SS} \\
N + N \rightleftharpoons N.N : & \qquad [N.N] = K_{\NN} [N]^2.       \label{K_NN} 
\end{eqnarray}

\noindent Surface hybridisation
\begin{eqnarray}
P + S \rightleftharpoons P.S : & \qquad [P.S] = K_{\PeS} [P][S].       \label{K_PS} \\
P + N \rightleftharpoons P.N : & \qquad [P.N] = K_{\PN} [P][N].       \label{K_PN} 
\end{eqnarray}

We begin by using Eqs.~(\ref{K_Sfold}) to (\ref{K_PN}) to eliminate concentrations of folded species and and of most duplex species from Eqs.~(\ref{xS}) to (\ref{thetaDef}).  From Eqs.~(\ref{pTotal}) and (\ref{thetaDef}) we obtain
\begin{eqnarray}
\theta_S = \frac{K_\PeS [S]}{(1 + K_\Pfold) + K_\PeS [S] + K_\PN [N]},   \label{thetaS1} \\
\theta_N = \frac{K_\PN [N]}{(1 + K_\Pfold) + K_\PeS [S] + K_\PN [N]},    \label{thetaN1}
\end{eqnarray}
and from Eqs.~(\ref{xS}) and (\ref{xN}), 
\begin{eqnarray}
x_S = \left(1 + K_\Sfold + K_\SN [N] \right)[S] + K_\SeS [S]^2 + [P.S], \\
x_N = \left(1 + K_\Nfold + K_\SN [S] \right)[N] + K_\NN [N]^2 + [P.N]. 
\end{eqnarray}
Following ref.~\cite{Binder09a} we make the reasonable assumptions
\begin{enumerate}
\item{$K_\SeS[S] << 1$: Specific targets will not easily encounter each other in bulk solution;}
\item{$K_\SN[S] << 1$: Very little of the depletion of nonspecific targets by bulk hybridisation is due to encounters with the specific targets in question;}
\item{$[P.N] <<x_N$: The proportion of nonspecific background depleted by hybridisation to the microarray is negligible.}
\end{enumerate}
With these assumptions, the above equations reduce to
\begin{eqnarray}
x_S = \left(1 + K_\Sfold + K_\SN [N] \right)[S] + [P.S], \label{xSreduced} \\
x_N = \left(1 + K_\Nfold \right)[N] + K_\NN [N]^2.   \label{xNreduced}
\end{eqnarray}
Eq.~(\ref{xNreduced}) is a quadratic in $[N]$ whose solution we will write as 
\begin{equation}
[N] = f_N(x_N, K_\NN, K_\Nfold).   \label{Nsoln}
\end{equation}
Previously (Eq.~(2.8) of~\cite{Binder06a} and Eq.~(1) of \cite{Binder09a}) the following approximation 
\begin{equation}
[N] \approx \frac{x_N}{1 + K_\Nfold + K_\NN x_N}  
\end{equation}
has been used, though this approximation is not necessary in the current context and is only included for comparison with previous work. We also have 
\begin{eqnarray}
[S] & = & \frac{x_S - [P.S]}{1 + K_\Sfold + K_\SN f_N(x_N, K_\NN, K_\Nfold)} \\ 
    & \approx & \frac{x_S - [P.S]}{1 + K_\Sfold + K_\SN x_N}, 
\end{eqnarray}
once again employing the same approximation.  

Substituting back into Eqs.~(\ref{thetaS1}) and (\ref{thetaN1}) gives 
\begin{eqnarray}
\theta_S = \frac{K_S \left(x_S - [P.S] \right)}{1 + X_N + K_S \left(x_S - [P.S] \right)},   \label{thetaS2} \\
\theta_N =  \frac{X_N}{1 + X_N + K_S \left(x_S - [P.S] \right)},   \label{thetaN2}  
\end{eqnarray}
where 
\begin{eqnarray}
X_N & = & \frac{K_\PN}{1 + K_\Pfold} f_N(x_N, K_\NN, K_\Nfold) \\
         & \approx & \frac{K_\PN x_N}{(1 + K_\Pfold)(1 + K_\Nfold + K_\NN x_N)},  \label{XNSapprox}
\end{eqnarray}
and
\begin{eqnarray}
K_S & = & \frac{K_\PeS}{(1 + K_\Pfold)(1 + K_\Sfold + K_\SN f_N(x_N, K_\NN, K_\Nfold))} \label{KSdefLocal1} \\
         & \approx & \frac{K_\PeS}{(1 + K_\Pfold)(1 + K_\Sfold + K_\SN x_N)}.  \label{KSdefLocal}
\end{eqnarray}
Finally, using Eq.~(\ref{thetaDef}), gives 
\begin{eqnarray}
\theta_S = \frac{K_S \left(x_S - p \theta_S \right)}{1 + X_N + K_S \left(x_S - p \theta_S \right)},   \label{thetaS3} \\
\theta_N =  \frac{X_N}{1 + X_N + K_S \left(x_S - p \theta_S \right)}.   \label{thetaN3}  
\end{eqnarray}
The quantity $X_N$ is known as the non-specific binding strength, and in the approximation of Eq.~(\ref{XNSapprox}) is often written in the form $X_N = K_N x_N$ where $K_N$ is an effective equilibrium constant for non-specific binding.  It is also common to define a specific binding strength $X_S = K_S x_S$ in terms of the effective specific equilibrium constant $K_S$ and specific target concentration.  

For given $x_S$, $x_N$, $p$ and equilibrium constants $K_r$, Eq.~(\ref{thetaS3}) is a quadratic in $\theta_S$ with a unique solution in $[0,1]$, namely 
\begin{equation}
\theta_S = \frac{1}{2}\left[ \frac{1 + X_N}{K_S p} + 1 + \frac{x_S}{p} - 
	\sqrt{\left( \frac{1 + X_N}{K_S p} + 1 + \frac{x_S}{p} \right)^2 - 4\frac{x_S}{p}} \right].
									\label{thetaSfromQuad}
\end{equation}
The required result is then
\begin{equation}
\theta = \theta_S + \theta_N = \frac{X_N + K_S \left(x_S - p \theta_S \right)}{1 + X_N + K_S \left(x_S - p \theta_S \right)} .               \label{thetaSoln}
\end{equation}
If post-hybridisation washing is significant, it is introduced into the model via specific and non-specific survival factors $w_S$ and $w_N$, where $1 >  w_S > w_N > 0$, giving 
\begin{equation}
\theta_{\rm after.wash} = w_S \theta_S + w_N \theta_N = \frac{w_S X_N + w_N K_S \left(x_S - p \theta_S \right)}{1 + X_N + K_S \left(x_S - p \theta_S \right)} .       
\end{equation}

\subsection{Global depletion}
\label{sec:AGlobalDepletion}

In the case of global depletion the target concentration specific to a given feature is assumed to be depleted by the hybridisation to all features which target the same chemical species.  Below we consider the case of a PM/MM pair of probe features, though the analysis readily generalises to any number of features addressing the same specific species.  We use superscripts PM and MM to indicate probe molecules on respective elements of a PM/MM pair, and denote by $S$ the target species complementary to the PM probe.  With these changes the set of input parameters become (1) the total specific target concentration
\begin{equation}
x_S = [S] + [S'] + [P^\PeM.S]+ [P^\MM.S] + [S.N] + [S.S],    \label{xSP}
\end{equation}
(2) an effective total non-specific target concentration 
\begin{equation}
x_N = [N] + [N'] + [P^\PeM.N] + [P^\MM.N] + [S.N] + [N.N],    \label{xNP}
\end{equation}
(3) an effective probe concentration, assumed to be the same for PM and MM, 
\begin{eqnarray}
p & = & [P^\PeM] + [{P^\PeM}'] + [P^\PeM.S] + [P^\PeM.N]    \nonumber \\
   & = & [P^\MM] + [{P^\MM}'] + [P^\MM.S] + [P^\MM.N],                        \label{pTotalP}
\end{eqnarray}
and a set of equilibrium constants $K^P_r$, which may or may not depend on $P = \PeM, \MM$, depending on the reaction $r$.  Our aim is now to determine a coverage fraction
\begin{equation}
\theta^P = \theta_S^P + \theta_N^P = \frac{[P^P.S]}{p} + \frac{[P^P.N]}{p},   \qquad P = \PeM, \MM \label{thetaDefP}
\end{equation}
for both elements of a probe pair.  

Analogous to Eqs.~(\ref{thetaS1}) and (\ref{thetaN1}) we have 
\begin{eqnarray}
\theta_S^P = \frac{K_\PeS^P [S]}{(1 + K_\Pfold^P) + K_\PeS^P [S] + K_\PN^P [N]},      \qquad P = \PeM, \MM\label{thetaSP1}, \\
\theta_N^P = \frac{K_\PN^P [N]}{(1 + K_\Pfold^P) + K_\PeS^P [S] + K_\PN^P [N]},       \qquad P = \PeM, \MM\label{thetaNP1}. 
\end{eqnarray}
After making the `reasonable assumptions' of the previous section, Eq.~(\ref{xSreduced}) becomes
\begin{equation}
x_S = \left(1 + K_\Sfold + K_\SN [N] \right)[S] + [P^\PeM.S]+ [P^\MM.S], \label{xSreducedP} \\
\end{equation}
and Eqs.(\ref{xNreduced}) and (\ref{Nsoln}) remain unchanged.  Then Eqs.~(\ref{thetaS2}) and (\ref{thetaN2}) become 
\begin{eqnarray}
\theta_S^P = \frac{K_S^P \left(x_S - [P^\PeM.S] - [P^\MM.S] \right)}
                    {1 + X_N^P + K_S^P \left(x_S - [P^\PeM.S] - [P^\MM.S]\right)},   \label{thetaSP2} \\
\theta_N^P =  \frac{X_N^P}{1 + X_N^P + K_S^P \left(x_S - [P^\PeM.S] - [P^\MM.S] \right)},   \label{thetaNP2}  
\end{eqnarray}
where 
\begin{equation}
X_N^P = \frac{K_\PN^P}{1 + K_\Pfold^P} f_N(x_N, K_\NN, K_\Nfold) ,  \label{XNPdef}
\end{equation}
and
\begin{equation}
K_S^P = \frac{K_\PeS^P}{(1 + K_\Pfold^P)(1 + K_\Sfold + K_\SN f_N(x_N, K_\NN, K_\Nfold))} . \label{KSPdef}
\end{equation}
Using Eq.~(\ref{thetaDefP}) then gives 
\begin{eqnarray}
\theta_S^P = \frac{K_S^P \left[x_S - p (\theta_S^\PeM +  \theta_S^\MM) \right]}
{1 + X_N^P + K_S^P \left[x_S - p (\theta_S^\PeM +  \theta_S^\MM) \right]},   \label{thetaSP3} \\
\theta_N^P =  \frac{X_N^P}
{1 + X_N^P + K_S^P \left[x_S - p (\theta_S^\PeM +  \theta_S^\MM) \right]}.   \label{thetaNP3} 
\end{eqnarray}
Summing Eq.~(\ref{thetaSP3}) over $P$ and defining $\theta_\Sum = \theta_S^\PeM +  \theta_S^\MM$, gives 
\begin{equation}
\theta_\Sum = \sum_{P = \PeM, \MM} 
\frac{K_S^P (x_S - p \theta_\Sum)}{1 + X_N^P + K_S^P (x_S - p \theta_\Sum)}.  \label{thetaSumP}
\end{equation}
This equation is cubic in $\theta_\Sum$, and can easily be solved numerically as a function of $x_S$, $K_S^P$, $X_N^P$ and $p$ using a Newton-Raphson algorithm.  The required coverage function is then 
\begin{equation}
\theta^P = \frac{X_N^P + K_S^P (x_S - p \theta_\Sum)}{1 + X_N^P + K_S^P (x_S - p \theta_\Sum)}, 
               \qquad P = \PeM, \MM.        \label{thetaSolnP}
\end{equation}
Again post-hybridisation can be introduced into the model via specific and non-specific survival factors, giving 
\begin{equation}
\theta^P_{\rm after.wash} = \frac{w_S^P X_N^P + w_N^P K_S^P (x_S - p \theta_\Sum)}{1 + X_N^P + K_S^P (x_S - p \theta_\Sum)}, 
               \qquad P = \PeM, \MM.        \label{postWashTheta}      
\end{equation}

\section{Analysis of the shape of the isotherms}
\label{sec:IsoShape}

We demonstrate that in the global PM/MM depletion model with $s = K^\PeM/K^\MM >1$ considered in Section~\ref{sec:ShapeIso}, the MM response curve acquires an inflection point for sufficiently high values of $K^\PeM$, while the PM response curve flattens without forming an inflection point as $K^\PeM$ increases.  

Defining $\phi(x) = K^\PeM(x - p\theta_\Sum)$, Eq.~(\ref{ThetaP}) gives $\Theta^\PeM = \phi/(1 + \phi)$ and $\Theta^\MM = \phi/(s + \phi)$, and thus 
\begin{equation}
\frac{d^2\Theta^\PeM}{dx^2} = \frac{\phi''}{(1 + \phi)^2 } - 2\frac{(\phi')^2}{(1 + \phi)^3 } , \qquad   
\frac{d^2\Theta^\MM}{dx^2} = \frac{s\phi''}{(s + \phi)^2 } - 2s\frac{(\phi')^2}{(s + \phi)^3 }, 
                                           \label{ThetaDerivatives} 
\end{equation}
while differentiating Eq.~(\ref{thetaSumDimsionless}) twice gives, 
\begin{equation}
-\frac{\phi''}{pK^\PeM} = \frac{\phi''}{(1 + \phi)^2 } - 2\frac{(\phi')^2}{(1 + \phi)^3 } + 
                                 \frac{s\phi''}{(s + \phi)^2 } - 2s\frac{(\phi')^2}{(s + \phi)^3 }. \label{phiSumDerivatives}      
\end{equation}
One easily checks that $\phi(0) = 0$, and thus Eq.~(\ref{phiSumDerivatives}) implies 
\begin{equation}
\phi''(0) = \frac{2(1 + s^2)}{s(1 + s + s/(pK^\PeM))} \phi'(0)^2. 
\end{equation}

Substituting back into Eq.(\ref{ThetaDerivatives}) at $x=0$ gives 
\begin{eqnarray}
\left. \frac{d^2\Theta^\PeM}{dx^2} \right|_{x=0} & = & \phi''(0) - 2 \phi'(0)^2 \nonumber \\
                 & = & 2\left( \frac{1 + s^2}{s(1 + s + s/(pK^\PeM))} - 1  \right)\phi'(0)^2 \nonumber \\
                 & < & 2\left( \frac{1 + s^2}{s(1 + s)} - 1  \right)\phi'(0)^2 \nonumber \\
                 & = & 2\frac{1 - s}{s(1 + s)} \phi'(0)^2 < 0, 
\end{eqnarray}
for $s>1$.  That is, the PM response curve is concave downwards at the origin for all physically relevant values of $s$.  In fact $d^2\Theta^\PeM/dx^2|_{x=0}$ increases from $-2(K^\PeM)^2$ at $p=0$ to $0$ as $p\rightarrow\infty$ and hence the response curve flattens to an almost straight line.  

Similarly we have 
\begin{eqnarray}
\left. \frac{d^2\Theta^\MM}{dx^2} \right|_{x=0} & = & \frac{1}{s}\phi''(0) - \frac{2}{s^2} \phi'(0)^2 \nonumber \\
                 & = & \frac{2}{s^2}\left( \frac{1 + s^2}{1 + s + s/(pK^\PeM)} - 1  \right)\phi'(0)^2, \nonumber \\
\end{eqnarray}
from which it follows that 
\begin{equation}
\left. \frac{d^2\Theta^\MM}{dx^2} \right|_{x=0} \lessgtr 0 
\qquad \mbox{according as} \qquad
pK^\PeM \lessgtr \frac{1}{s - 1}.
\end{equation}
Thus, the MM response curve has an inflection point for $pK^\PeM > 1/(s - 1)$.  

\section{Critique of Li et al.}
\label{sec:LiCritique}

We point out errors in the thermodynamic model proposed in a recent paper by Li et al.~\cite{Li08}.  The primary source of error in this paper is an incorrect use of the law of mass action in Eq.~(3) of their paper describing the rate $\dot{n}_{\rm in}$ of binding of specific and non-specific targets to probes.  In the notation of Li et al., the corrected form of the equation is 
\begin{equation}
\frac{\dot{n}_{\rm in}}{N_A V} = (1 - \alpha - \beta)pk_b ([T] + [N]),       \label{corrected3}
\end{equation}  
where $\alpha$ and $\beta$ are specific and non-specfic coverage fractions (equivalent to our $\theta_S$ and $\theta_N$), $p$ is the effective probe concentration, $[T]$ and $[N]$ free specific and non-specific target concentrations,  $k_b$ the reaction rate for binding (assumed to be determined by a rate-determining initiation step and therefore the same for specific and non-specific targets), $N_A$ is avogadro's number and $V$ volume of the hybridisation solution.  The factor $([T] + [N])$ is missing from Li et al.'s paper, either intentionally or through an oversight, but must be present if the reaction proceeds at a rate proportional to the product of the concentrations of each of the reactants.  

With this correction, Eq.~(5) of ref.~\cite{Li08} balancing the forward and backward reaction rates becomes 
\begin{equation}
(1 - \alpha - \beta)pk_b ([T] + [N]) = \alpha pk_d + \beta pk_n, 
\end{equation}
where $k_d$ and $k_n$ are dissociation rate constants for specific and non-specific duplexes respectively.  Eq.~(8) of ref.~\cite{Li08} is best derived by balancing the forward and backward rates for specific and nonspecific targets separately: 
\begin{eqnarray}
(1 - \alpha - \beta)pk_b [T] & = & \frac{\dot{n}_{\rm in}^{(T)}}{N_A V}= 
                                                          \frac{\dot{n}_{\rm out}^{(T)}}{N_A V} = \alpha pk_d, \nonumber \\
(1 - \alpha - \beta)pk_b [N] & = & \frac{\dot{n}_{\rm in}^{(N)}}{N_A V}= 
                                                          \frac{\dot{n}_{\rm out}^{(N)}}{N_A V} = \alpha pk_n, \nonumber 
\end{eqnarray}
giving 
\begin{equation}
\beta = \frac{k_d[N]}{k_n[T]}\alpha, \nonumber
\end{equation}
in agreement with Eq.~(8) of ref.~\cite{Li08}.  In fact this equation cannot be derived without the assumption that the forward reactions are driven at rates proportional to the target concentrations, as used in Eq.~(\ref{corrected3}) above, but not in Eq.~(3) of ref.~\cite{Li08}.  Substituting this back into Eq.~(\ref{corrected3}) gives the corrected form of Eq.~(9) of ref.~\cite{Li08}, 
\begin{equation}
\alpha = \frac{1}{1 + (k_d/k_n)([N]/[T]) + (k_d/k_b)(1/[T])} = \frac{K_T[T]}{1 + K_T[T] + K_N[N]}, \label{correctedAlpha}
\end{equation}
where we define specific and non-specific equilibrium constants $K_T = k_b/k_d$ and $K_N = k_b/k_n$ respectively.  A similar calculation gives the non-specfic coverage fraction as 
\begin{equation}
\beta = \frac{K_N[N]}{1 + K_T[T] + K_N[N]}.  \label{correctedBeta}
\end{equation}

Li et al.\ incorporate target depletion by hybridisation from the supernatant solution by making the substitution $[T] = \hat{T} - \alpha p$, where $\hat{T}$ is the nominal spike-in concentration.  With appropriate changes of notation, the corrected equations~(\ref{correctedAlpha}) and (\ref{correctedBeta}) with this substitution are essentially nothing more than simplified versions of the Ono model~\cite{Ono08}, or of our local depletion model Eqs.~(\ref{thetaS3}) and (\ref{thetaN3}), without inclusion of probe or target folding or bulk hybridisation in the supernatant solution.  Li et al.\ then proceed to fit their model to the U133 Affymetrix latin square data set.  However, the above substitution corresponds to local, not global, depletion, which we have demonstrated in Section~\ref{AffyDataAnalysis} is not appropriate for this data set.  

Finally we note that Li et al.'s Eq.~(12) for the specific coverage fraction (the corrected form of which is Eq.~(\ref{correctedAlpha})), namely
\begin{equation}
\alpha = \frac{1}{1 + k_d[1/k_b + \gamma/(\hat{T} - \alpha p)]} \qquad \mbox{[{\em sic}]},
\end{equation}
where $\gamma = (1/k_n + 1/k_b)[N]$, cannot be correct by the following reasoning.  In the absence of a non-specific complex background ($[N] \rightarrow 0$, and thus $\gamma \rightarrow 0$), this equation predicts that the coverage fraction should be independent of spike-in concentration ($\alpha \rightarrow 1/(1 + k_d/k_b)$), and indeed equal to their predicted binding affinity dependent saturation coverage over the whole range of spike-in concentrations $\hat{T}$.  This is obviously wrong, as evidenced by a version of Affymetrix's U95a latin square spike-in experiment without complex background~\cite{Burden08} in which the experimentally obtained coverage fraction clearly responds to target concentration.  

%%%%%%%%%%%%%%%%%%%%%%%%%%%%%%%%%%%%%%%%%%%
\section*{Glossary}
\begin{description}
\item[{\it Hybridisation.}]{The reversible chemical reaction by which target molecules in solution bind to probes attached to the microarray surface to form duplexes.}  
\item[{\it Microarray.}]{A high-throughput device for detecting the presence of large biological molecules (DNA, RNA or proteins) of specific known letter sequences via their binding to molecules of complementary sequences attached to a solid surface.  They are high-throughput in the sense that large numbers of sequences are tested for in a single device.  The microarrays discussed here are oligonucleotide gene expression microarrays, that is, they have short DNA probes and are intended for the detection of expressed genes through their messenger RNA.}  
\item[{\it Non-specific hybridisation.}]{The hybridisation of target molecules with sequences other than those of the intended sequence.  When dealing with microarrays with a PM/MM (perfect match/mismatch) design, `non-specific' is used to mean `non-PM-specific', that is, hybridisation of target molecules which are not complementary to the PM sequence, irrespective of whether they are binding to the PM or MM member of a probe pair.}
\item[{\it Perfect match/Mismatch probes.}]{(Conventionally abbreviated as PM and MM.)  A common design in Affymetrix GeneChip microarrays is to represent each targeted nucleotide sequence by two neighbouring probe features: the PM, whose DNA probe sequence is exactly complementary to the target sequence, and the MM, whose DNA sequence is identical to the PM sequence except that the base in the central position of the probe sequence is replaced by a base complementary to that in the PM sequence.  The idea behind the MMs is that they should respond to non-specific targets in a way similar to their PM partner, and can be used as a way of controlling biases due to non-specific hybridisation.}
\item[{\it Probe.}]{A biological molecule attached to the microarray surface during fabrication.}
\item[{\it Spike-in experiment.}]{An experiment in which known concentrations of a specific set of target molecules are artificially added to a solution not otherwise containing those specific targets, and the solution hybridised onto microarrays.}
\item[{\it Target.}]{A biological molecule in the solution hybridised onto the microarray during a laboratory experiment.}  
\end{description}
%%%%%%%%%%%%%%%%%%%%%%%%%%%%%%%%%%%%%%%%%%%
\pagebreak

\end{document}